\pdfoutput=1
\documentclass[twocolumn]{aastex63}

\footnoterule
\usepackage{amsmath}
\usepackage{xfrac}
\begin{document}

\title{ALMA Observations of Proper Motions of the Dust Clumps in the Protoplanetary Disk MWC~758}

\author{Kuo, I-Hsuan Genevieve}
\affiliation{Academia Sinica Institute of Astronomy and Astrophysics, 11F of Astro-Math Bldg, 1, Sec. 4, Roosevelt Rd, Taipei 10617, Taiwan}
\affiliation{Department of Astronomy and Steward Observatory, University of Arizona, Tucson, AZ 85721, USA}

\author{Yen, Hsi-Wei}
\affiliation{Academia Sinica Institute of Astronomy and Astrophysics, 11F of Astro-Math Bldg, 1, Sec. 4, Roosevelt Rd, Taipei 10617, Taiwan}

\author{Gu, Pin-Gao}
\affiliation{Academia Sinica Institute of Astronomy and Astrophysics, 11F of Astro-Math Bldg, 1, Sec. 4, Roosevelt Rd, Taipei 10617, Taiwan}


\begin{abstract}

To study the dust dynamics in the dust trapping vortices in the  protoplanetary disk around MWC~758, we analyzed the 1.3 mm continuum images of the MWC~758 disk obtained with the Atacama Large Millimeter/submillimeter Array (ALMA) in 2017 and 2021. We detect proper motions of 22 mas and 24 mas in the two dust clumps at radii of 0\farcs32 and 0\farcs54 in the disk on the plane of the sky, respectively. On the assumption that the dust clumps are located in the disk midplane, the velocities of the observed proper motions along the azimuthal direction of the inner and outer dust clumps are sub- and super-Keplerian, respectively, and both have angular velocities corresponding to the Keplerian angular velocity at a radius of $0\farcs46\pm0\farcs04$. This deviation from the Keplerian motion is not expected in the conventional theory of vortices formed by the Rossby wave instability. The observed non-Keplerian proper motions of the dust clumps are unlikely due to the disk warp and eccentricity, nor be associated with any predicted planets. The two dust clumps are likely spatially coincident with the infrared spirals. In addition, we detect the changes in the intensity profiles of the dust clumps over the four-year span. Therefore, we suggest that the observed proper motions are possibly due to changes in the density distributions in the dust clumps caused by their interaction with the spirals in the disk. 
\end{abstract}

\section{Introduction} \label{sec:intro}

Crescents, or asymmetric dust intensity enhancements have been identified in a number of protoplanetary disks such as MWC~758, HD~142527, HD~135344B, HD~143006 \citep{Bae_PPVII}. Crescents have been widely postulated to be vortices based on analysis of their morphological properties \citep[e.g.][]{Caz_2018, Yen_2020, Bae_PPVII}. A vortex is a potential vorticity minimum as a result of the Rossby wave instability (RWI) at the edges of gaps or rings \citep{Lovelace_1999, Li_2001, Ono_2016}.  
Axisymmetric dust traps at radial pressure maximum in disks have been known to stall the rapid radial drift of dust towards the central star \citep{Weiden_1977}. Vortices further lead to asymmetric dust trapping azimuthally \citep{Birnstiel_2013}. These dust trapping mechanisms have critical implications to planet formation as they can raise the local dust-to-gas ratio or trigger local gravitational instability \citep[e.g.][]{BargenSommeria, Johansen_2007, Pinilla_2012a, Pinilla_2012b}, enhancing grain growth and the possibility for particles to coagulate into planetesimal size \citep{ChiangnYoudin, Lyra_2024}.

However, the hypothesis that the crescents observed in dust continuum are dust trapped by vortices has not yet been confirmed because of the difficulty in resolving the vorticity and line-of-sight velocity residuals on the order of few percent of Keplerian velocity \citep{Robert_2020}. There are also alternate explanations to these asymmetric continuum structures, including spiral density waves \citep{vanderMarel_2016}, temperature variations due to shadowing effects \citep{Marino_2015_shadow}, lumps at the edge of circumbinary cavities \citep{Miranda_2017}, or dust trapping at Lagrangian points of a planetary companion \citep{Mont_2020}. Different formation mechanisms of the crescents can be distinguished through their morphology and motion relative to the disk. For instance, vortices propagate at their local Keplerian speed, while spiral density waves corotate with the perturber, and the rotation period of the shadows is dependent on the cooling time of the disk and precession period of the inner disk. Thus, observations revealing motions of crescents can shed light on the nature of the asymmetric intensity enhancements in dusty disks.


MWC~758 is an Herbig Ae star with a stellar mass of 1.5--2 $M_\odot$ at a distance of 160 pc \citep{Garufi_2018, Vioque_2018, Isella_2010, Kuo_2022}. It is surrounded by a protoplanetary disk exhibiting complex structures in its dust morphology \citep{Dong_2018}. It hosts a large inner cavity with a radius of $\sim50$ au, an eccentric dust ring, spirals observed in both infrared and  millimeter wavelengths, and two dust crescents \citep{Grady_2013, Benisty_2015, Ren_2018, Dong_2018}. Simulations have shown that RWI-generated vortices in MWC~758 can trap dust particles and reproduce the morphology of the crescents seen in continuum observations of MWC~758 \citep{Baruteau_2019}. In this paper, we present the ALMA 1.3 mm continuum image of MWC~758 taken in 2017 and 2021, and our objective is to constrain the proper motion of the crescents in MWC~758 over the span of two observational epochs and study their nature.

\section{Data and Calibration} \label{sec:data}
\begin{figure*}[ht!]
\epsscale{1.15}
\plotone{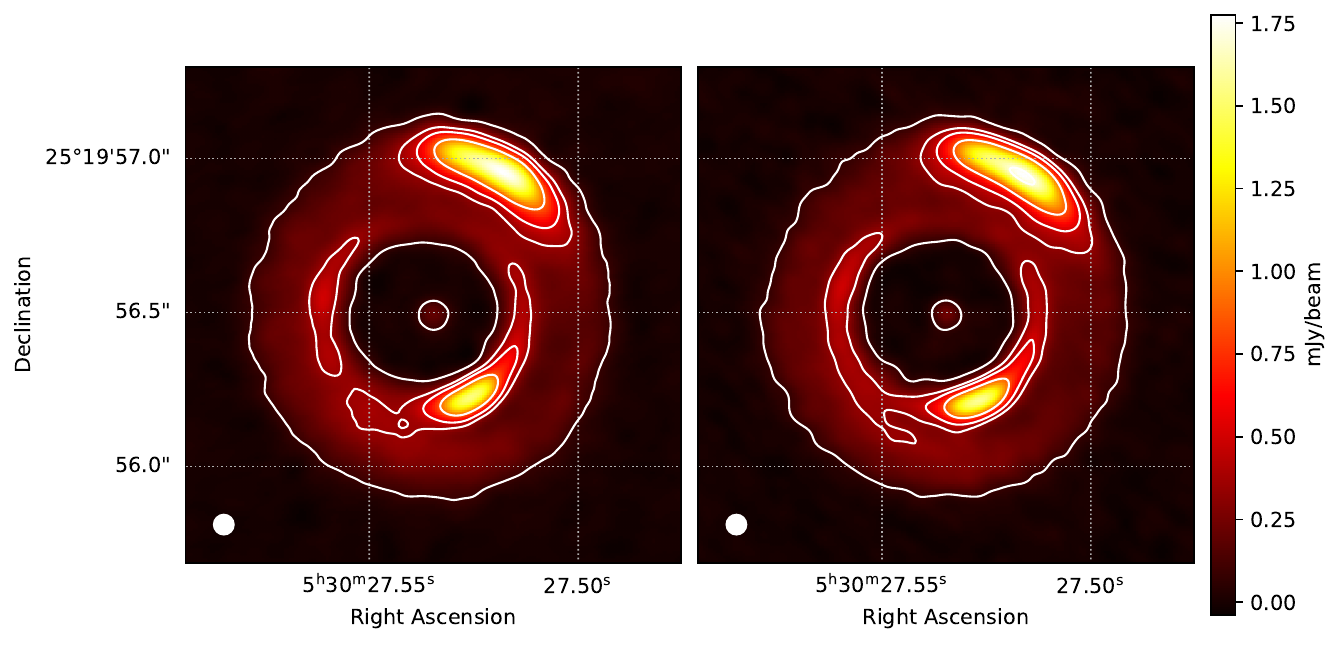}
\caption{ALMA 1.3 mm continuum maps of the MWC~758 disk from the 2017 (left) and (right) 2021 data. The two maps have the same beam size of 0\farcs071, shown as white ellipses at the bottom left corners, and are generated with similar uv coverages. The contour levels are 5$\sigma$, 20$\sigma$, 30$\sigma$, 50$\sigma$, and 100$\sigma$. \label{fig:datamap}}
\end{figure*}

 In this paper, we make use of two ALMA Band 6 datasets, one from our Cycle 8 project 2021.1.01356.S, and another retrieved from the ALMA public archive under the Cycle 5 project 2017.1.00940.S. Our observations of MWC~758 were executed on 2021 October 24, 27, and 29. All eight execution blocks (EBs) were observed with the array configuration C-8. The combined data set has projected baseline lengths ranging from 46.8 m to 8.9 km. 
 The spectral setup consists of four spectral windows. One spectral window with a bandwidth of 2 GHz was assigned to the continuum emission centered at 232.499 GHz. $\mathrm{CO \ (2-1)}$, $^{13}\mathrm{CO \ (2-1)}$, and $\mathrm{C}^{18}\mathrm{O \ (2-1)}$ lines were also observed simultaneously at high spectral resolutions of 30.518 kHz.  
 \par
 
 The 2017.1.00940.S archival data was observed on October 10, 11, 15, 16, and December 9, 17, 27, 28 in 2017. Four EBs were observed with the configuration C-10, with projected baseline lengths ranging between 41.4 m and 16.2 km. Six other EBs were observed with the configuration C-6, with projected baseline lengths ranging between 15.1 m and 3.1 km.
 The spectral setup consists of four spectral windows. Two spectral windows each with a bandwidth of 2 GHz were assigned to the continuum emission centered at 234.191 GHz and 217.499 GHz, respectively. $\mathrm{CO \ (2-1)}$, $^{13}\mathrm{CO \ (2-1)}$, and $\mathrm{C}^{18}\mathrm{O \ (2-1)}$ lines were also observed simultaneously. 
The line data in this project has been presented in \citet{Wolfer_2023}.
\par

In the present paper, we focus on the continuum data. 
We calibrated all the raw visibility data with the pipeline scripts using the Common Astronomy Software Applications (CASA) of version 6.2.1.7 for the 2021.1.01356.S data and version 5.1.1-5 for the 2017.1.00940.S data.
We first performed self calibration on the phase of the continuum data and generated continuum images for individual EBs using the CASA task {\tt tclean}. The inner disk around the star located inside the cavity are detected in all EBs. 

To account for the proper motion of the star and the astrometric accuracy of the ALMA observations\footnote{https://almascience.nrao.edu/proposing/technical-handbook}, we adopted this inner disk as a reference source to align the data from different EBs. We assumed that its peak position coincident with the stellar position, and confirmed that its morphology is consistent between the two epochs (Appendix \ref{app:inner}). 
We fitted a two-dimensional Gaussian function to the inner disk and measured its center. Then we aligned the data of the EBs from the same projects with this inner disk center using the CASA tasks {\tt phaseshift} and {\tt fixplanets}, and the data from each project were concatenated. 
\par

Next, we performed self calibration on the phase of the concatenated continuum data for each project, and then performed one round of self calibration on the amplitude and phase. We generated continuum images using the self-calibrated, concatenated data for each project, measured the center of the inner disk again, and aligned the two datasets taken in 2017 and 2021. 
The two datasets were centered at ($\alpha_\mathrm{J2000}$, $\delta_\mathrm{J2000}$) = (05:30:27.5347, +25:19:56.492).
The uncertainty in the measured inner disk center and thus in aligning the 2017 and 2021 datasets is 2 mas.

After the self calibration and alignment of the datasets, we generated final continuum images using {\tt tclean} with the same baseline range to achieve a similar uv coverage for both datasets.
The pixel size was adopted to be 8 mas. Finally, we convolved the two images to the same circular beam size of $0\farcs071\times0\farcs071$. 
The signal-to-noise ratios of the final continuum images were improved by 30\% and 40\% for the 2017 and 2021 data with the self-calibration, respectively.
The rms noise is 10.5 $\mu$Jy in the 2017 image and is 9.4 $\mu$Jy in the 2021 image. 
In addition, we have confirmed that the difference in the uv coverages and observing frequencies of the 2017 and 2021 data does not cause any false detection of proper motion (Appendix \ref{app:uv}). 

\section{Analysis and Results}

\begin{figure*}[ht!]
\epsscale{1.15}
\plotone{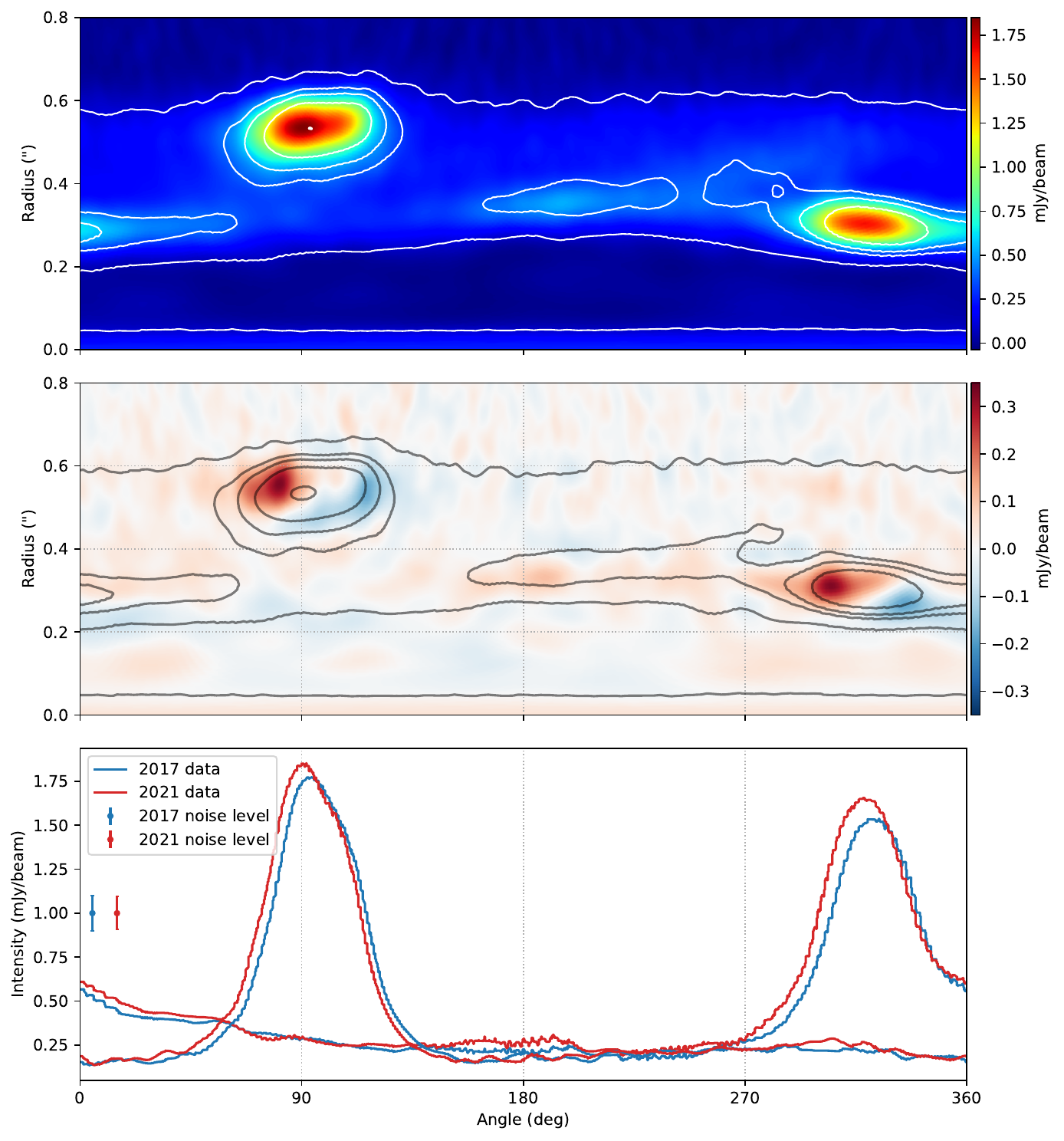}
\caption{Azimuthal Intensity Distribution and Displacement. (\textit{upper panel}) ALMA 1.3 mm continuum maps deprojected in the polar coordinates. The azimuthal angle is defined counter-clockwise from the minor axis of the disk. The color scale and contours present the 2021 and 2017 data, respectively. The contour levels are 5$\sigma$, 20$\sigma$, 30$\sigma$, 50$\sigma$, and 100$\sigma$. (\textit{middle panel}) Intensity residuals (color) after subtracting the 2021 from 2017 maps presented in the upper panel. The contours present the 2021 continuum map with the contour levels the same as those in the upper panel. (\textit{lower panel}) Azimuthal intensity profiles of the inner and outer clumps passing through the continuum peaks of the clumps extracted from the 2017 (blue) and 2021 (red) continuum maps. The inner clump is centered around 322$\arcdeg$, and the outer clump at 95$\arcdeg$. 
\label{fig:polmap}}
\end{figure*}

\subsection{1.3 mm Continuum Images in 2017 and 2021}

\par 
The image of the dust continuum from both epochs are shown in Figure \ref{fig:datamap}. The qualities of these two images are similar. The total intensities of the MWC~758 disk measured in these two images are 5.3 Jy and 5.1 Jy with uncertainties of 20 $\mu$Jy and 40 $\mu$Jy, respectively. The total fluxes measured in the two epochs are consistent within 4\%. This flux difference is smaller than the 5\% absolute flux accuracy of ALMA observations in Band 6. The noise levels of the two images of 10.5 $\mu$Jy and 9.4 $\mu$Jy are also consistent within 10\%. 

In both images, the eccentric dust ring is clearly detected, and the two known, prominent dust clumps are observed to the northwest and southeast of the disk at high signal-to-noise ratios (S/N) of $>100$ at the peaks. The north (outer) clump is brighter and more radially extended than the south (inner) clump. The overall shapes and intensity distributions of the two clumps in the two epochs are similar by comparing the contours of the 2017 and 2021 images.
Below, we measure the positions of these prominent structures and investigate their relative offsets between two epochs. 


\subsection{Proper Motion of Eccentric Dust Ring}
The inner edge of the dust cavity is known to be eccentric with an eccentricity of $e=0.10\pm0.01$ and a phase angle of its semi major axis of $275\arcdeg\pm10\arcdeg$ \citep{Dong_2018}.
We fitted an eccentric ring model $r=\frac{a(1-e^2)}{1-e\cos{f}}$ to the $5\sigma$ contours delineating the inner edge of the dust cavity in the deprojected 2017 and 2021 images, where $r$ is the radial polar coordinate centered at one focus, $a$ is the semi-major axis of the ellipse, and $f$ is the polar angle from the pericenter. 
We measured the phase angle ($\theta$) and $a$ of the elliptical dust cavity at 5$\sigma$ in both epochs to be 
($\theta_{2017}=285\arcdeg\pm11\arcdeg$, $a_{2017}=0\farcs24\pm0\farcs06$) and ($\theta_{2021}=277\arcdeg\pm15\arcdeg$, $a_{2021}=0\farcs23\pm0\farcs06$), respectively.
The difference in the semi major axis and its phase angle is within the $1\sigma$ uncertainty, so no proper motion of the eccentric dust ring is observed. 

\subsection{Proper Motion of Dust Clumps}

\subsubsection{Displacement of Peak Position in Image}
We measured the peak positions of the two dust clumps in the 2017 and 2021 images on the plane of sky by fitting a parabola to 3 by 3 pixels centered at the maximum pixel in the clumps, since their intensity distributions are curved and are not Gaussian-like. The right ascension (RA) and declination (Dec) offsets of the peak of the northern clump with respect to the image center are measured to be ($-$225.0 mas, 461.8 mas) in 2017 and ($-$247.1 mas, 452.5 mas) in 2021, and those of the southern clump are measured to be ($-$122.6 mas, $-$266.0 mas) in 2017 and ($-$104.5 mas, $-$278.6 mas) in 2021. The accuracy in determining peak positions is approximately the angular resolution divided by S/N. With our high angular resolution of 71 mas and S/N $>$ 100, the position accuracy is better than 1 mas, so the uncertainty in the relative offsets between the peak positions measured in 2017 and 2021 is limited by the accuracy in aligning the two images and is 2 mas. 

Our results reveal the proper motions of the two dense clumps over a time span of four years. 
The proper motions along RA and Dec of the northern clump are ($-$22 mas, $-$9 mas), and those of the southern clump are (18 mas, $-$13 mas) at a high significance of 5--10$\sigma$. 
Assuming an inclination angle of $16\arcdeg$ and PA of the disk major axis of $240\arcdeg$ \citep{Kuo_2022}, 
the proper motions in the radial and azimuthal directions of the northern clump at a radius of 0\farcs54 are derived to be ($-$2 mas, 2.4$\arcdeg$ or 23 mas),
and those of the southern clump at a radius of 0\farcs32 to be (6 mas, 4.2$\arcdeg$ or 22 mas), in the disk plane.
The proper motions of the dust clumps are primarily along the azimuthal direction in the disk plane, and the directions of these proper motions are consistent with the direction of the disk rotation. 
\par

To further investigate time variations of the positions and intensity distributions of the clumps, we deprojected both images with an inclination angle of $16\arcdeg$ and PA of the disk major axis of $240\arcdeg$ \citep{Kuo_2022}, and transformed the images into polar coordinates as shown in the upper panel of Figure \ref{fig:polmap}. Then we subtracted the 2021 image from the 2017 image in the polar coordinates (Fig. \ref{fig:polmap} middle panel). Systematic residuals at a high significance of 20$\sigma$ are observed in the clumps. No significant residuals are detected outside the clumps. 
There are gradients of the residuals in the clumps, which are due to their proper motions.
The residual gradients are larger along the azimuthal direction than the radial direction because the proper motions of the dust clumps are primarily in the azimuthal direction. 
Because the overall intensity distributions of the dust clumps are similar between the two epochs and the azimuthal proper motions are dominant, below we measure the proper motions of the dust clumps by comparing their azimuthal and radial intensity profiles between the two epochs instead of only their peak positions .

\begin{figure*}[ht!] 
\epsscale{1.15}
\plotone{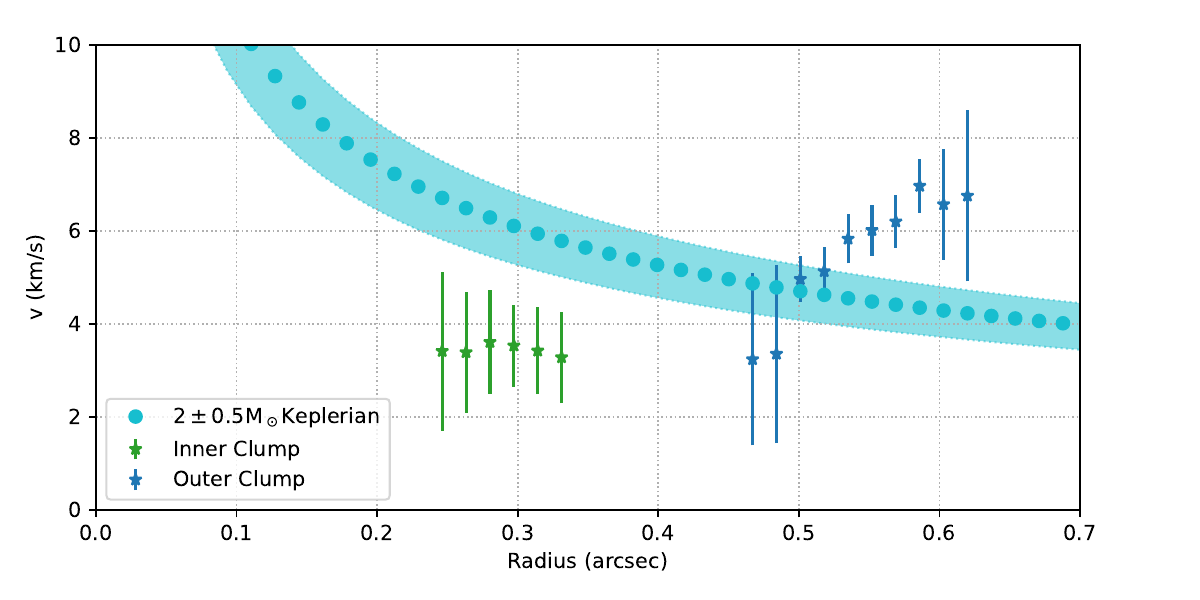}
\caption{Measured azimuthal velocity of the proper motions of the inner and outer clumps (green and blue data points, respectively,) in comparison with the Keplerian rotational profile of MWC~758, a $2\ M_\odot$ star (cyan dots) with uncertainty of $\pm0.5\ M_\odot$ (shading). The error bars show the 1$\sigma$ uncertainties of the measured azimuthal velocities. The inner clump peaks at a radius of $0\farcs32$ and azimuthal angle of $320\arcdeg$, while the outer clump peaks at a radius of $0\farcs53$ and azimuthal angle of $90\arcdeg$.} 
\label{fig:azv}
\end{figure*}

\par

\subsubsection{Proper Motion in Azimuthal Direction from Deprojected Image} \label{sec:az_method}
To constrain the proper motion of the clumps as a whole by taking into account the entire intensity distribution instead of just the peak position, we extracted intensity profiles of the dust clumps along the azimuthal direction at a series of radii in steps of one-third of the beam size from the 2017 and 2021 images in the polar coordinates. We have confirmed that the exact step size to bin the intensity profiles radially does not affect the result.
The lower panel in Figure \ref{fig:polmap} presents an example of the azimuthal intensity profiles at the radii of the intensity peaks of the dust clumps. The dust clumps are distinct from the rest of the disk and have intensities above 20$\sigma$, and exhibit similar intensity profiles in 2017 and 2021 with the 2021 peak intensity brighter by 4.5\%. 

The observed intensity profiles of the dust clumps are not Gaussian-like and are asymmetric with respect to the peaks. We measured the azimuthal proper motion of the dust clumps in the disk plane from the relative offsets between the entire intensity profiles of the dust clumps observed in two epochs.
We shifted the azimuthal intensity profile at a given radius from the 2017 image along the azimuthal direction and calculated the residuals after subtracting the shifted 2017 profile from the 2021 profile at the same radius. We estimated the azimuthal proper motion of the dust clumps at the given radius by determining the azimuthal displacement that minimizes the chi-squared statistic $\chi^2$. We estimate the uncertainty of our measurements using the confidence limits at $1\sigma$ significance that is defined by $\chi^2_{min}+1$, where $\chi^2_{min}$ is the minimal chi-squared. The alignment uncertainty of 2 mas was also considered by systematically shifting the images by 2 mas, and we confirmed that our result is not affected by this alignment uncertainty.

The azimuthal proper motions in the dust clumps at radii of $0\farcs$24--$0\farcs35$ and $0\farcs46$--$0\farcs63$ estimated with this analysis are converted to angular and azimuthal velocities and are shown in Fig.~\ref{fig:azv}. Beyond these radial ranges, we could not find any displacements along the azimuthal direction to further reduce the residuals after subtracting the 2017 from 2021 profiles. 
After applying the displacements and minimizing the residuals, the gradients of the residual along the azimuthal direction are mostly removed, and only the gradients along the radial direction are clearly seen in the dust clumps in the residual map (Appendix \ref{app:rad}).

The spectroscopic mass of MWC 758 has been estimated to be $1.5M_\odot$ and $1.9M_\odot$ with stellar evolutionary models \citep{Garufi_2018, Vioque_2018}, and the dynamical mass from Keplerian fitting is $2M_\odot$ \citep{Isella_2010,Kuo_2022}. Thus in Figure \ref{fig:azv} we plot Keplerian velocity with a central star mass $2M_\odot\pm0.5M_\odot$ to compare with the measured proper motions. The measured azimuthal velocity from the proper motion of the inner dust clump is slower than the Keplerian velocity by $3\pm1$ km s$^{-1}$ (or 50\%), while that of the outer dust clumps is faster than the Keplerian velocity by $1.5\pm0.6$ km s$^{-1}$ (or 33\%). The two dust clumps have comparable angular velocities of 4.48$\times$$10^{-10}$ s$^{-1}$ or 0.8 degrees per year, corresponding to the Keplerian angular velocity at radii of $0\farcs46\pm0\farcs04$, considering the uncertainty in the stellar mass (Fig.~\ref{fig:azv}).  
Changing the position and inclination angles within 5 degrees for deprojection, which are their uncertainties reported in the literature, the changes in the measured velocity of the proper motion is less than 0.1 km/s. Thus, the result is not affected by the uncertainty in the disk orientation.



\subsubsection{Proper Motion in Radial Direction from Deprojected Image}

We applied the same approach and extracted the intensity profiles along the radial direction at given azimuthal angles. 
An example of the radial intensity profiles at the azimuthal angles of the intensity peaks of the dust clumps is presented in Appendix \ref{app:rad}. We compared these radial intensity profiles between the two epochs and determined the radial displacements that can minimize the residuals after subtracting one from the other. Although there are offsets of 2--4 mas between the peak positions of the 2017 and 2021 radial intensity profiles, we did not find any radial displacement that further minimize the residuals, suggesting that there is no detectable proper motion along the radial direction of the dust clumps as a whole. 

The radial gradients of the residuals in the dust clumps are due to the minor variations at a level of 0.1--0.2 mJy in the intensity distributions along the radial direction instead of proper motions. 
This variation corresponds to 10--20$\sigma$, and thus it is unlikely caused by the difference in the uv coverages between the two data sets (Appendix \ref{app:uv}).
In both dust clumps, their radial intensity profiles of the inner and outer parts of the clumps become fainter and brighter in 2021, respectively (Appendix \ref{app:rad}). 
As the two dust clumps are located at different position angles, this intensity variation along the radial direction is unlikely due to any error in aligning the two images.


\section{Discussion}


By analyzing the 1.3 mm continuum images obtained with ALMA in 2017 and 2021, we detect proper motion of 24 mas and 22 mas in the northern and southern dust clumps in the MWC~758 disk on the plane of the sky, and our position accuracy is 2 mas. We do not detect any proper motion of its eccentric dust ring (or cavity). 
The non-detection of proper motion of the eccentric dust ring is consistent with the expectation that the precession timescale of the eccentric ring excited by a putative planet is substantially longer than the observational timescale of four years \citep{Hsieh_2012}. 


Assuming that the two dust clumps are located in the disk midplane having the same position angle and inclination as the outer disk, we converted the measured azimuthal proper motions into azimuthal velocities (Section \ref{sec:az_method}), and found that the inner and outer clumps move at sub-Keplerian and super-Keplerian speeds by 50\% and 33\% at a significance of the $3\sigma$ level, respectively (Figure \ref{fig:azv} upper panel). These two dust clumps have been suggested to be dust-trapping vortices \citep{Marino_2015, Baruteau_2019, Casassus_2019}, while our measured proper motions are not consistent with the expected Keplerian motion from the conventional vortex theory 
\citep{Ono_2016}. 

We note that the MWC~758 inner and outer disks are misaligned by $50\arcdeg$ \citep{Francis_2020}, which may hint that the outer disk itself can be also warped. By assuming that the motion of the dust clumps follows the circular Keplerian rotation in a warped disk, we have found no solution for a combination of disk inclination and position angle that could reproduce the measured proper motion (Appendix \ref{app:warp}), and thus a warped outer disk (if present) cannot explain the non-Keplerian motion of the dust clumps. In addition, the eccentricity of the outer disk is $\sim$0.1 at a radius of 0\farcs32 and decreases outwards \citep{Dong_2018}. Therefore, the deviation from the circular Keplerian rotation due to the eccentric motion is at most 10\% of the Keplerian velocity \citep{Kuo_2022}. The non-Keplerian motion of the dust clumps cannot be caused by the disk eccentricity. 

The angular velocities of both dust clumps correspond to the angular Keplerian velocity at a radius of $0\farcs46\pm0\farcs04$ (Figure \ref{fig:azv}). If the non-Keplerian motion of the dust clumps is caused by a perturber, then the location of the perturber likely lies at this radius, the corotating radius. However, this radius matches none of the predicted locations of the putative planets so far, which are at $\lesssim$0\farcs2 or $\gtrsim$0\farcs6 \citep{Baruteau_2019, Calcino_2020, Ren_2020, Wagner_2023}. Thus, the non-Keplerian motion of the dust clumps may not be associated with putative planets in the disk. 

In the infrared scattered-light images, two prominent spirals moving at a pattern speed of $0\fdg22\pm0\fdg03\ \mathrm{yr}^{-1}$ have been observed, and they likely overlap with the two dust clumps detected in the millimeter continuum \citep{Dong_2018, Ren_2020}.
In the theoretical simulations, dust grains can be trapped by the pressure bump of a spiral, and the density distribution of dust structures can be deformed when they cross the spiral \citep{Price_2018}.
Our analysis also detects changes in the intensity profiles of the dust clumps. 
Furthermore, in addition to the eccentric gas rotation at the radius of the eccentric dust ring, the ALMA $^{13}\mathrm{CO}$ observations have also tentatively revealed non-Keplerian gas motion coincident with the locations of the dust clumps observed in the 1.3 mm continuum \citep{Kuo_2022}.
These observational results may hint at interaction between the spirals and dust clumps in the MWC~758 disk.

Such interaction with the spirals can deform the density distributions in the dust clumps, leading to changes in the intensity profiles and peak positions in the 1.3 mm continuum images. In this case, even though the vortices are moving as a whole at the Keplerian velocity in the MWC~758 disk, the proper motions of the dust clumps observed in the millimeter continuum emission can appear to be super or sub-Kelperian. For example, if the spiral is crossing the northern dust clump and has passed its vortex center in 2017, the dust density can be enhanced at the tail of the dust clump, where the spiral and dust clump intersect, compared to the center of the vortex. The observed displacement between the continuum peak positions between 2017 and 2021 would appear to be super-Keplerian motion as the intensity profile shifts back toward the vortex center. Future disk modeling is needed to elucidate the scenario.

The dust distribution in a vortex being deformed by a spiral passing through has been tentatively seen in hydrodynamical simulations of MWC~758 \citep{Baruteau_2019}. Therefore, the interaction between the spirals and dust clumps deforming the density distributions in the dust clumps may cause the observed non-Keplerian proper motion. High-resolution molecular-line observations revealing the gas velocity in the dust clumps are essential to investigate possible interaction between the spirals and dust clumps and its impact on the dust density distributions. Besides, in this scenario, we expect to observe different proper motions in the dust continuum at different wavelengths. 
In the outer dust clump in the northwest in the MWC~758 disk, larger dust grains observed at 9 mm are found to be better trapped in the vortex, compared to smaller dust grains observed at 1.3 mm \citep{Casassus_2019}.
Thus, these larger dust grains are expected to more closely follow the motion of the vortex and move at the Keplerian velocity, while the distribution of the smaller dust grains are more easily affected by the interaction with the spiral. Future disk modeling and measurements of proper motions in the dust continuum at longer wavelengths, such as 9 mm for comparison with the previous VLA observations \citep{Casassus_2019}, can test this hypothesis.

\section{Summary}\label{sec:summary}
To study the dust dynamics in dust trapping vortices in the MWC~758 disk, we analyzed its 1.3 mm continuum images taken with ALMA in 2017 and 2021 at an angular resolution of 0\farcs07. We detect proper motions of the intensity peaks of the inner and outer dust clumps in the MWC~758 disk, to be 22 mas and 24 mas on the plane of the sky, respectively, while there is no detection of proper motion of its eccentric ring. The proper motions of the dust clumps are primarily along the azimuthal direction, and are in the same direction as the disk rotation. 

We have deprojected and compared the intensity profiles of the dust clumps observed in 2017 and 2021, on the assumption that the dust clumps are in the disk midplane, and measured the azimuthal velocities of the dust clumps from their displacements over the four-year span. The outer and inner dust clumps at radii of 0\farcs54 and 0\farcs32 move at super- and sub-Keplerian velocities with deviations up to 50\% of the Keplerian velocities, respectively. Their angular velocities correspond to the Keplerian angular velocity at a radius of 0\farcs46. In addition, we detect the variations in the radial intensity profiles of the dust clumps. 

We explored several potential causes for the non-Keplerian proper motions of the dust clumps. The observed non-Keplerian proper motions cannot be explained with the disk warp or eccentricity, and are unlikely associated with the predicted planetary perturbers. The dust clumps likely spatially overlap with the infrared spiral, and show variations in their continuum intensity distributions. Therefore, the observed non-Keplerian proper motions of the dust clumps are possibly due to their interactions with the spiral density waves. Such interaction can deform the dust distributions within the clumps, leading to the observed deviations from Keplerian motion, while the vortices as a whole propagates at the Keplerian velocity. Future measurements of proper motions in dust continuum at longer wavelengths are essential to further test this scenario.



\acknowledgments

We thank Clement Baruteau for fruitful discussions on the theories and simulation results of vortices and dust-gas interaction. 
This paper makes use of the following ALMA data: ADS/JAO.ALMA\#2021.1.01356.S and ADS/JAO.ALMA\#2017.1.00940.S.. ALMA is a partnership of ESO (representing its member states), NSF (USA) and NINS (Japan), together with NRC (Canada), NSTC and ASIAA (Taiwan), and KASI (Republic of Korea), in cooperation with the Republic of Chile. The Joint ALMA Observatory is operated by ESO, AUI/NRAO and NAOJ.
H.-W.Y.\ acknowledges support from the National Science and Technology Council (NSTC) in Taiwan through grant NSTC 113-2112-M-001-035- and 113-2124-M-001-008- and from the Academia Sinica Career Development Award (AS-CDA-111-M03).

\bibliography{main}

\begin{thebibliography}{}
\expandafter\ifx\csname natexlab\endcsname\relax\def\natexlab#1{#1}\fi
\providecommand{\url}[1]{\href{#1}{#1}}
\providecommand{\dodoi}[1]{doi:~\href{http://doi.org/#1}{\nolinkurl{#1}}}
\providecommand{\doeprint}[1]{\href{http://ascl.net/#1}{\nolinkurl{http://ascl.net/#1}}}
\providecommand{\doarXiv}[1]{\href{https://arxiv.org/abs/#1}{\nolinkurl{https://arxiv.org/abs/#1}}}

\bibitem[{{Bae} {et~al.}(2023){Bae}, {Isella}, {Zhu}, {Martin}, {Okuzumi}, \& {Suriano}}]{Bae_PPVII}
{Bae}, J., {Isella}, A., {Zhu}, Z., {et~al.} 2023, in Astronomical Society of the Pacific Conference Series, Vol. 534, Protostars and Planets VII, ed. S.~{Inutsuka}, Y.~{Aikawa}, T.~{Muto}, K.~{Tomida}, \& M.~{Tamura}, 423, \dodoi{10.48550/arXiv.2210.13314}

\bibitem[{{Barge} \& {Sommeria}(1995)}]{BargenSommeria}
{Barge}, P., \& {Sommeria}, J. 1995, \aap, 295, L1, \dodoi{10.48550/arXiv.astro-ph/9501050}

\bibitem[{Baruteau {et~al.}(2019)Baruteau, Barraza, Pérez, Casassus, Dong, Lyra, Marino, Christiaens, Zhu, Carmona, Debras, \& Alarcon}]{Baruteau_2019}
Baruteau, C., Barraza, M., Pérez, S., {et~al.} 2019, Monthly Notices of the Royal Astronomical Society, 486, 304, \dodoi{10.1093/mnras/stz802}

\bibitem[{{Benisty} {et~al.}(2015){Benisty}, {Juhasz}, {Boccaletti}, {Avenhaus}, {Milli}, {Thalmann}, {Dominik}, {Pinilla}, {Buenzli}, {Pohl}, {Beuzit}, {Birnstiel}, {de Boer}, {Bonnefoy}, {Chauvin}, {Christiaens}, {Garufi}, {Grady}, {Henning}, {Huelamo}, {Isella}, {Langlois}, {M{\'e}nard}, {Mouillet}, {Olofsson}, {Pantin}, {Pinte}, \& {Pueyo}}]{Benisty_2015}
{Benisty}, M., {Juhasz}, A., {Boccaletti}, A., {et~al.} 2015, \aap, 578, L6, \dodoi{10.1051/0004-6361/201526011}

\bibitem[{{Birnstiel} {et~al.}(2013){Birnstiel}, {Dullemond}, \& {Pinilla}}]{Birnstiel_2013}
{Birnstiel}, T., {Dullemond}, C.~P., \& {Pinilla}, P. 2013, \aap, 550, L8, \dodoi{10.1051/0004-6361/201220847}

\bibitem[{{Calcino} {et~al.}(2020){Calcino}, {Christiaens}, {Price}, {Pinte}, {Davis}, {van der Marel}, \& {Cuello}}]{Calcino_2020}
{Calcino}, J., {Christiaens}, V., {Price}, D.~J., {et~al.} 2020, \mnras, 498, 639, \dodoi{10.1093/mnras/staa2468}

\bibitem[{{Casassus} {et~al.}(2019){Casassus}, {Marino}, {Lyra}, {Baruteau}, {Vidal}, {Wootten}, {P{\'e}rez}, {Alarcon}, {Barraza}, {C{\'a}rcamo}, {Dong}, {Sierra}, {Zhu}, {Ricci}, {Christiaens}, \& {Cieza}}]{Casassus_2019}
{Casassus}, S., {Marino}, S., {Lyra}, W., {et~al.} 2019, \mnras, 483, 3278, \dodoi{10.1093/mnras/sty3269}

\bibitem[{{Cazzoletti} {et~al.}(2018){Cazzoletti}, {van Dishoeck}, {Pinilla}, {Tazzari}, {Facchini}, {van der Marel}, {Benisty}, {Garufi}, \& {P{\'e}rez}}]{Caz_2018}
{Cazzoletti}, P., {van Dishoeck}, E.~F., {Pinilla}, P., {et~al.} 2018, \aap, 619, A161, \dodoi{10.1051/0004-6361/201834006}

\bibitem[{{Chiang} \& {Youdin}(2010)}]{ChiangnYoudin}
{Chiang}, E., \& {Youdin}, A.~N. 2010, Annual Review of Earth and Planetary Sciences, 38, 493, \dodoi{10.1146/annurev-earth-040809-152513}

\bibitem[{{Dong} {et~al.}(2018){Dong}, {Liu}, {Eisner}, {Andrews}, {Fung}, {Zhu}, {Chiang}, {Hashimoto}, {Liu}, {Casassus}, {Esposito}, {Hasegawa}, {Muto}, {Pavlyuchenkov}, {Wilner}, {Akiyama}, {Tamura}, \& {Wisniewski}}]{Dong_2018}
{Dong}, R., {Liu}, S.-y., {Eisner}, J., {et~al.} 2018, \apj, 860, 124, \dodoi{10.3847/1538-4357/aac6cb}

\bibitem[{Francis \& van~der Marel(2020)}]{Francis_2020}
Francis, L., \& van~der Marel, N. 2020, The Astrophysical Journal, 892, 111, \dodoi{10.3847/1538-4357/ab7b63}

\bibitem[{{Garufi} {et~al.}(2018){Garufi}, {Benisty}, {Pinilla}, {Tazzari}, {Dominik}, {Ginski}, {Henning}, {Kral}, {Langlois}, {M{\'e}nard}, {Stolker}, {Szulagyi}, {Villenave}, \& {van der Plas}}]{Garufi_2018}
{Garufi}, A., {Benisty}, M., {Pinilla}, P., {et~al.} 2018, \aap, 620, A94, \dodoi{10.1051/0004-6361/201833872}

\bibitem[{{Grady} {et~al.}(2013){Grady}, {Muto}, {Hashimoto}, {Fukagawa}, {Currie}, {Biller}, {Thalmann}, {Sitko}, {Russell}, {Wisniewski}, {Dong}, {Kwon}, {Sai}, {Hornbeck}, {Schneider}, {Hines}, {Moro Mart{\'\i}n}, {Feldt}, {Henning}, {Pott}, {Bonnefoy}, {Bouwman}, {Lacour}, {Mueller}, {Juh{\'a}sz}, {Crida}, {Chauvin}, {Andrews}, {Wilner}, {Kraus}, {Dahm}, {Robitaille}, {Jang-Condell}, {Abe}, {Akiyama}, {Brandner}, {Brandt}, {Carson}, {Egner}, {Follette}, {Goto}, {Guyon}, {Hayano}, {Hayashi}, {Hayashi}, {Hodapp}, {Ishii}, {Iye}, {Janson}, {Kandori}, {Knapp}, {Kudo}, {Kusakabe}, {Kuzuhara}, {Mayama}, {McElwain}, {Matsuo}, {Miyama}, {Morino}, {Nishimura}, {Pyo}, {Serabyn}, {Suto}, {Suzuki}, {Takami}, {Takato}, {Terada}, {Tomono}, {Turner}, {Watanabe}, {Yamada}, {Takami}, {Usuda}, \& {Tamura}}]{Grady_2013}
{Grady}, C.~A., {Muto}, T., {Hashimoto}, J., {et~al.} 2013, \apj, 762, 48, \dodoi{10.1088/0004-637X/762/1/48}

\bibitem[{Hsieh \& Gu(2012)}]{Hsieh_2012}
Hsieh, H.-F., \& Gu, P.-G. 2012, The Astrophysical Journal, 760, 119, \dodoi{10.1088/0004-637x/760/2/119}

\bibitem[{{Isella} {et~al.}(2010){Isella}, {Natta}, {Wilner}, {Carpenter}, \& {Testi}}]{Isella_2010}
{Isella}, A., {Natta}, A., {Wilner}, D., {Carpenter}, J.~M., \& {Testi}, L. 2010, \apj, 725, 1735, \dodoi{10.1088/0004-637X/725/2/1735}

\bibitem[{{Johansen} {et~al.}(2007){Johansen}, {Oishi}, {Mac Low}, {Klahr}, {Henning}, \& {Youdin}}]{Johansen_2007}
{Johansen}, A., {Oishi}, J.~S., {Mac Low}, M.-M., {et~al.} 2007, \nat, 448, 1022, \dodoi{10.1038/nature06086}

\bibitem[{{Kuo} {et~al.}(2022){Kuo}, {Yen}, {Gu}, \& {Chang}}]{Kuo_2022}
{Kuo}, I. H.~G., {Yen}, H.-W., {Gu}, P.-G., \& {Chang}, T.-E. 2022, \apj, 938, 50, \dodoi{10.3847/1538-4357/ac9228}

\bibitem[{{Li} {et~al.}(2001){Li}, {Colgate}, {Wendroff}, \& {Liska}}]{Li_2001}
{Li}, H., {Colgate}, S.~A., {Wendroff}, B., \& {Liska}, R. 2001, \apj, 551, 874, \dodoi{10.1086/320241}

\bibitem[{{Lovelace} {et~al.}(1999){Lovelace}, {Li}, {Colgate}, \& {Nelson}}]{Lovelace_1999}
{Lovelace}, R.~V.~E., {Li}, H., {Colgate}, S.~A., \& {Nelson}, A.~F. 1999, \apj, 513, 805, \dodoi{10.1086/306900}

\bibitem[{{Lyra} {et~al.}(2024){Lyra}, {Yang}, {Simon}, {Umurhan}, \& {Youdin}}]{Lyra_2024}
{Lyra}, W., {Yang}, C.-C., {Simon}, J.~B., {Umurhan}, O.~M., \& {Youdin}, A.~N. 2024, \apjl, 970, L19, \dodoi{10.3847/2041-8213/ad5af6}

\bibitem[{{Marino} {et~al.}(2015{\natexlab{a}}){Marino}, {Perez}, \& {Casassus}}]{Marino_2015_shadow}
{Marino}, S., {Perez}, S., \& {Casassus}, S. 2015{\natexlab{a}}, \apjl, 798, L44, \dodoi{10.1088/2041-8205/798/2/L44}

\bibitem[{{Marino} {et~al.}(2015{\natexlab{b}}){Marino}, {Perez}, \& {Casassus}}]{Marino_2015}
---. 2015{\natexlab{b}}, \apjl, 798, L44, \dodoi{10.1088/2041-8205/798/2/L44}

\bibitem[{{Miranda} {et~al.}(2017){Miranda}, {Mu{\~n}oz}, \& {Lai}}]{Miranda_2017}
{Miranda}, R., {Mu{\~n}oz}, D.~J., \& {Lai}, D. 2017, \mnras, 466, 1170, \dodoi{10.1093/mnras/stw3189}

\bibitem[{{Montesinos} {et~al.}(2020){Montesinos}, {Garrido-Deutelmoser}, {Olofsson}, {Giuppone}, {Cuadra}, {Bayo}, {Sucerquia}, \& {Cuello}}]{Mont_2020}
{Montesinos}, M., {Garrido-Deutelmoser}, J., {Olofsson}, J., {et~al.} 2020, \aap, 642, A224, \dodoi{10.1051/0004-6361/202038758}

\bibitem[{{Ono} {et~al.}(2016){Ono}, {Muto}, {Takeuchi}, \& {Nomura}}]{Ono_2016}
{Ono}, T., {Muto}, T., {Takeuchi}, T., \& {Nomura}, H. 2016, \apj, 823, 84, \dodoi{10.3847/0004-637X/823/2/84}

\bibitem[{{Pinilla} {et~al.}(2012{\natexlab{a}}){Pinilla}, {Benisty}, \& {Birnstiel}}]{Pinilla_2012a}
{Pinilla}, P., {Benisty}, M., \& {Birnstiel}, T. 2012{\natexlab{a}}, \aap, 545, A81, \dodoi{10.1051/0004-6361/201219315}

\bibitem[{{Pinilla} {et~al.}(2012{\natexlab{b}}){Pinilla}, {Birnstiel}, {Ricci}, {Dullemond}, {Uribe}, {Testi}, \& {Natta}}]{Pinilla_2012b}
{Pinilla}, P., {Birnstiel}, T., {Ricci}, L., {et~al.} 2012{\natexlab{b}}, \aap, 538, A114, \dodoi{10.1051/0004-6361/201118204}

\bibitem[{{Price} {et~al.}(2018){Price}, {Cuello}, {Pinte}, {Mentiplay}, {Casassus}, {Christiaens}, {Kennedy}, {Cuadra}, {Sebastian Perez}, {Marino}, {Armitage}, {Zurlo}, {Juhasz}, {Ragusa}, {Laibe}, \& {Lodato}}]{Price_2018}
{Price}, D.~J., {Cuello}, N., {Pinte}, C., {et~al.} 2018, \mnras, 477, 1270, \dodoi{10.1093/mnras/sty647}

\bibitem[{{Ren} {et~al.}(2018){Ren}, {Dong}, {Esposito}, {Pueyo}, {Debes}, {Poteet}, {Choquet}, {Benisty}, {Chiang}, {Grady}, {Hines}, {Schneider}, \& {Soummer}}]{Ren_2018}
{Ren}, B., {Dong}, R., {Esposito}, T.~M., {et~al.} 2018, \apjl, 857, L9, \dodoi{10.3847/2041-8213/aab7f5}

\bibitem[{{Ren} {et~al.}(2020){Ren}, {Dong}, {van Holstein}, {Ruffio}, {Calvin}, {Girard}, {Benisty}, {Boccaletti}, {Esposito}, {Choquet}, {Mawet}, {Pueyo}, {Stolker}, {Chiang}, {Boer}, {Debes}, {Garufi}, {Grady}, {Hines}, {Maire}, {M{\'e}nard}, {Millar-Blanchaer}, {Perrin}, {Poteet}, \& {Schneider}}]{Ren_2020}
{Ren}, B., {Dong}, R., {van Holstein}, R.~G., {et~al.} 2020, \apjl, 898, L38, \dodoi{10.3847/2041-8213/aba43e}

\bibitem[{{Robert} {et~al.}(2020){Robert}, {M{\'e}heut}, \& {M{\'e}nard}}]{Robert_2020}
{Robert}, C.~M.~T., {M{\'e}heut}, H., \& {M{\'e}nard}, F. 2020, \aap, 641, A128, \dodoi{10.1051/0004-6361/201937414}

\bibitem[{{van der Marel} {et~al.}(2016){van der Marel}, {Cazzoletti}, {Pinilla}, \& {Garufi}}]{vanderMarel_2016}
{van der Marel}, N., {Cazzoletti}, P., {Pinilla}, P., \& {Garufi}, A. 2016, \apj, 832, 178, \dodoi{10.3847/0004-637X/832/2/178}

\bibitem[{{Vioque} {et~al.}(2018){Vioque}, {Oudmaijer}, {Baines}, {Mendigut{\'\i}a}, \& {P{\'e}rez-Mart{\'\i}nez}}]{Vioque_2018}
{Vioque}, M., {Oudmaijer}, R.~D., {Baines}, D., {Mendigut{\'\i}a}, I., \& {P{\'e}rez-Mart{\'\i}nez}, R. 2018, \aap, 620, A128, \dodoi{10.1051/0004-6361/201832870}

\bibitem[{{Wagner} {et~al.}(2023){Wagner}, {Stone}, {Skemer}, {Ertel}, {Dong}, {Apai}, {Spalding}, {Leisenring}, {Sitko}, {Kratter}, {Barman}, {Marley}, {Miles}, {Boccaletti}, {Assani}, {Bayyari}, {Uyama}, {Woodward}, {Hinz}, {Briesemeister}, {Lawson}, {M{\'e}nard}, {Pantin}, {Russell}, {Skrutskie}, \& {Wisniewski}}]{Wagner_2023}
{Wagner}, K., {Stone}, J., {Skemer}, A., {et~al.} 2023, Nature Astronomy, 7, 1208, \dodoi{10.1038/s41550-023-02028-3}

\bibitem[{{Weidenschilling}(1977)}]{Weiden_1977}
{Weidenschilling}, S.~J. 1977, \mnras, 180, 57, \dodoi{10.1093/mnras/180.2.57}

\bibitem[{{W{\"o}lfer} {et~al.}(2023){W{\"o}lfer}, {Facchini}, {van der Marel}, {van Dishoeck}, {Benisty}, {Bohn}, {Francis}, {Izquierdo}, \& {Teague}}]{Wolfer_2023}
{W{\"o}lfer}, L., {Facchini}, S., {van der Marel}, N., {et~al.} 2023, \aap, 670, A154, \dodoi{10.1051/0004-6361/202243601}

\bibitem[{{Yen} \& {Gu}(2020)}]{Yen_2020}
{Yen}, H.-W., \& {Gu}, P.-G. 2020, \apj, 905, 89, \dodoi{10.3847/1538-4357/abc55a}

\end{thebibliography}
\bibliographystyle{aasjournal}

\begin{appendix}

\section{Use of Inner Disk for Image Alignment}\label{app:inner}

\begin{figure*}[ht!]
\epsscale{1.15}
\plotone{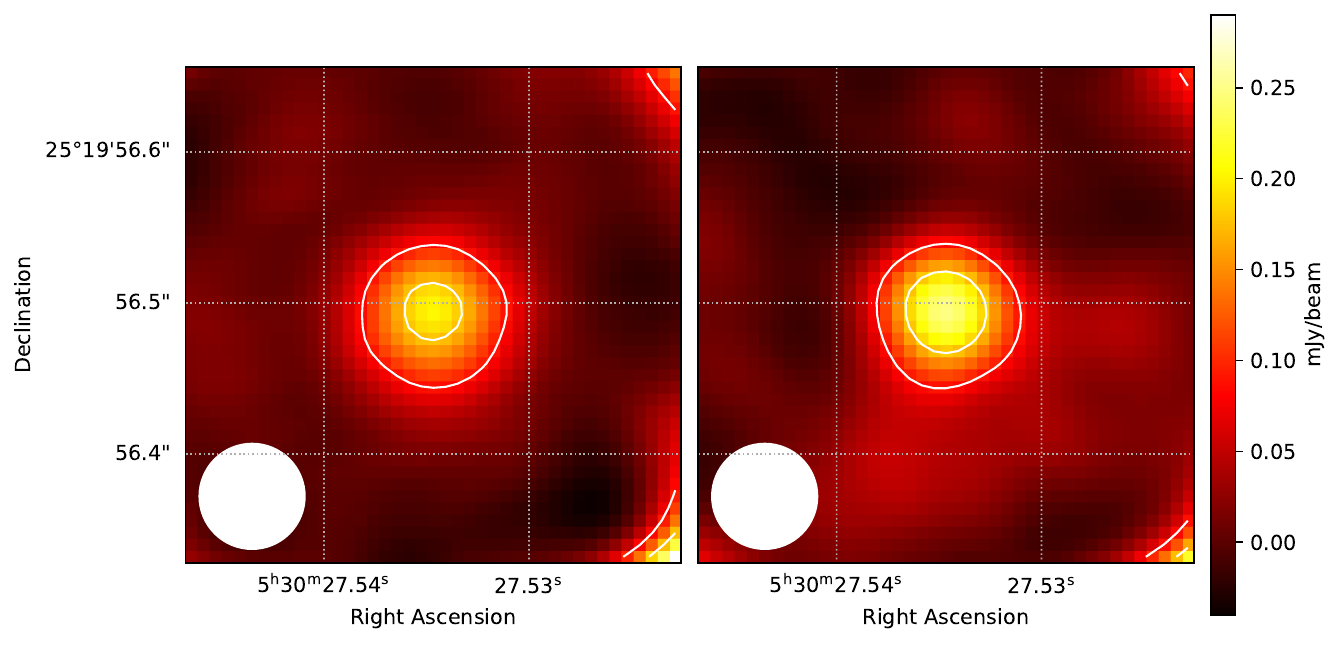}
\caption{ALMA 1.3 mm continuum maps of the inner disk of MWC~758 from the 2017 (left) and (right) 2021 data. The two maps have the same beam size of 0\farcs071, shown as white ellipses at the bottom left corners, and are generated with similar uv coverages. The contour levels are 5$\sigma$, 10$\sigma$. \label{fig:zoom}}
\end{figure*}

Figure \ref{fig:zoom} presents a zoom-in to the inner disks of MWC~758 observed in 2017 and 2021. The inner disk is marginally resolved by the beam size of 71 mas, and is isolated from other disk structures. Its overall shape does not undergo significant change over the course of 4 years.

\section{Continuum Intensity Distribution in the Radial Direction} \label{app:rad}

\begin{figure*}[ht!] 
\epsscale{1.15}
\plotone{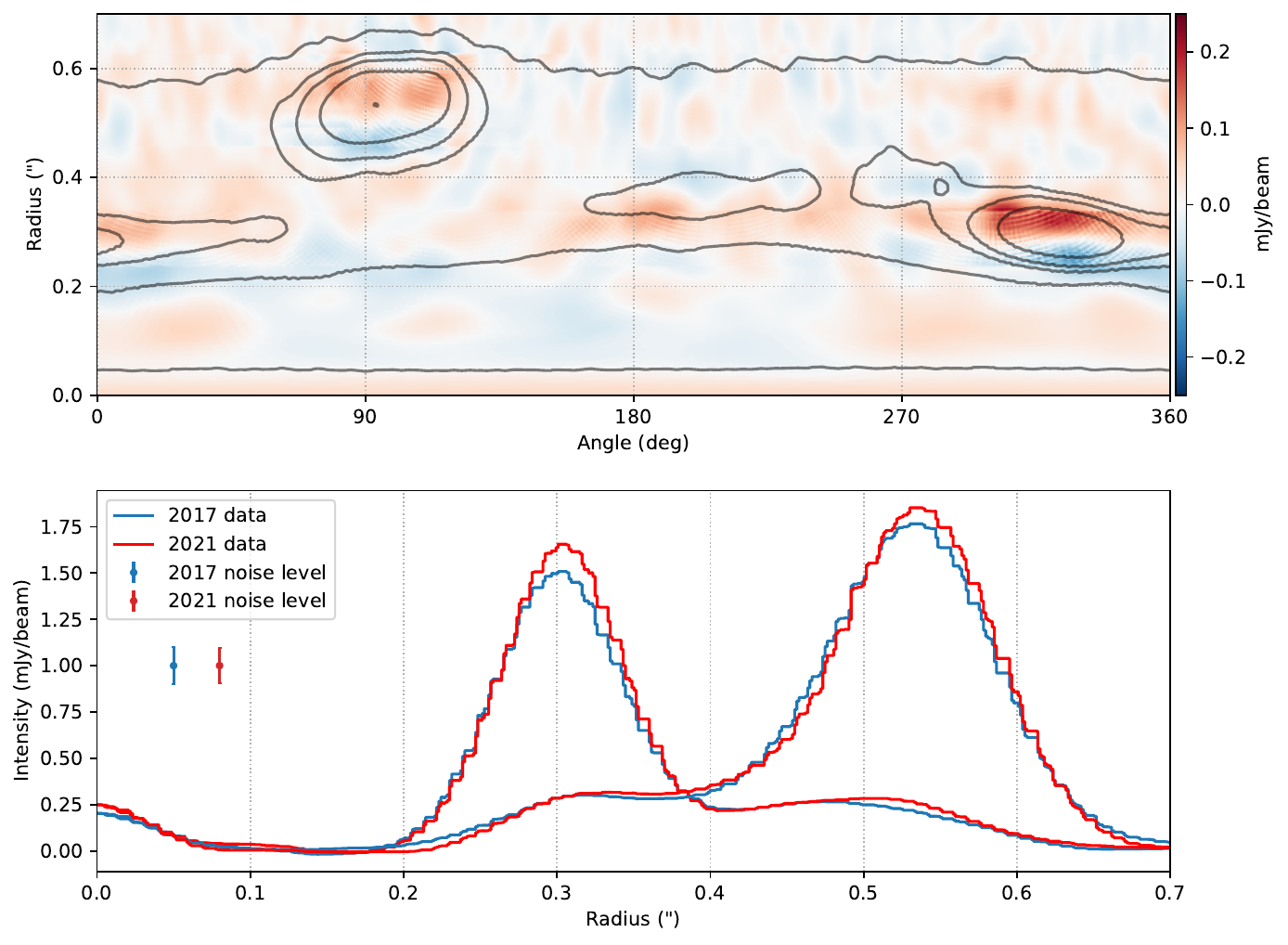}
\caption{Radial Intensity Distribution and Displacement. (\textit{upper panel}) Intensity residuals (color) after correcting for the measured azimuthal proper motions and subtracting the 2017 from 2021 maps in the polar coordinates. The contours present the 2017 continuum map with the contour levels the same as those in Figure \ref{fig:polmap}. (\textit{lower panel}) \label{fig:radial} Radial intensity profiles of the inner and outer clumps passing through the continuum peaks of the clumps extracted from the 2017 (blue) and 2021 (red) continuum maps after correcting for the measured azimuthal proper motions. }
\end{figure*}

Figure \ref{fig:radial} presents the intensity residual map between the 2017 and 2021 data after applying the measured azimuthal displacement in Section \ref{sec:az_method} and the radial intensity profiles of the inner and outer clump at azimuthal angles of 320$\arcdeg$ and 90$\arcdeg$, respectively.

\section{Difference between UV coverage and observing frequency in 2017 and 2021}\label{app:uv}
In order to validate that the observed changes in the continuum intensity distributions do not arise from the difference in the uv coverages between the two datasets, we adopted the clean model generated by the CASA task {\tt tclean} from the 2021 data as our model image, sampled it with the uv coverages of the 2017 and 2021 data using the CASA task {\tt ft}, and made model visibility data. Then we generated simulated images from our model visibility data, so the two simulated images have the same model input but different uv coverages. We subtracted one simulated image from the other, and the residuals are on the order of the noise level of the observations. We repeated the same procedure by adopting the clean model from the 2017 data, and the results were the same. Therefore, we do not expect any false detection of proper motion due to the difference in the uv coverages between the 2017 and 2021 data. 
\par
We also generated continuum images centered at 233GHz and 218GHz from the 2017 data, and conducted the analysis on these two images. There is no proper motion detected between these two images, and thus the wavelength difference does not effect our results.


\section{Projection effects due to possible disk warp} \label{app:warp}
We explored all possible disk orientations. For a given disk orientation, we deprojected the peak positions of the dust clumps in 2017, and derived its radius in the disk midplane. Then we calculated its displacement following the Keplerian rotation around MWC~758 over the four year span, and predicted its peak position in 2021. Finally, we projected this predicted peak position in the disk midplane onto the plane of the sky, and compared it with the observed peak position in 2021. In these calculations, we could not find a disk orientation that can match the predicted and observed peak positions in 2021. Therefore, the observed non-Keplerian proper motions of the dust clumps cannot be due to Keplerian rotation on a different disk plane.  
\section{Angular Velocity of the Dust Clumps}

\begin{figure*}[ht!] 
\epsscale{1.15}
\plotone{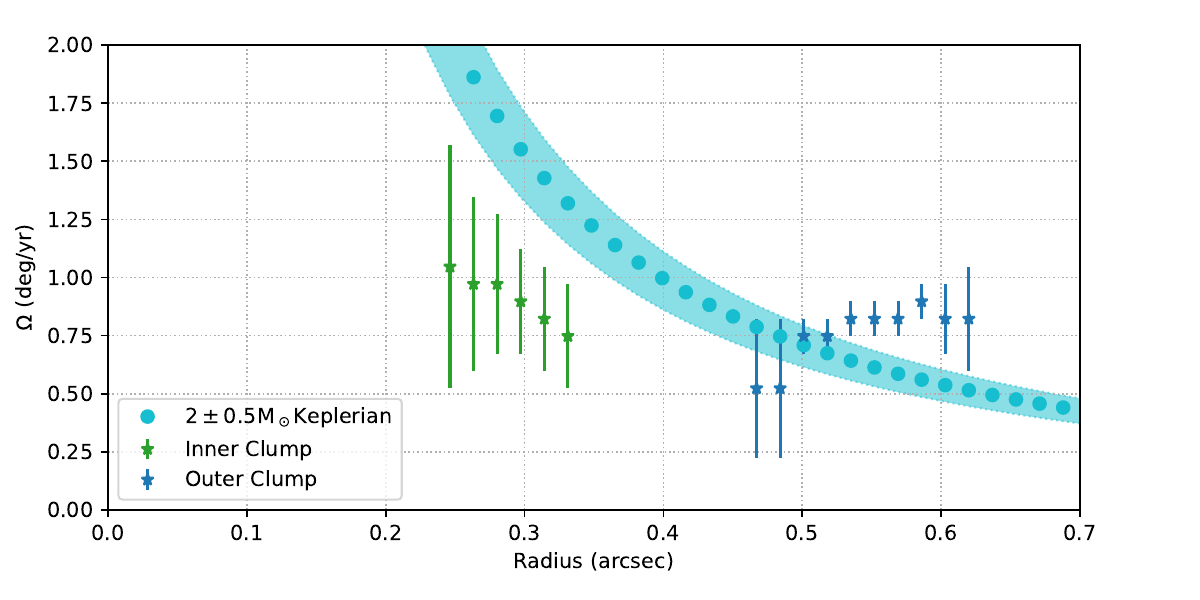}
\caption{Same as the Figure 3 but for the angular velocity. \label{fig:omega}}
\end{figure*}

Figure \ref{fig:omega} presents the measured angular velocity of the clumps at different radii, in units of degrees per year for comparison with pattern speeds of other disk structures.

\end{appendix}

\end{document}